\documentclass[onecolumn]{svjour3}                     
\smartqed  
\usepackage{graphicx}
\usepackage{amsmath,amssymb}
\usepackage{subfigure}
\usepackage{setspace}

\begin{document}

\title{Radio Resource Allocation Algorithms for Multi-Service OFDMA
Networks: The Uniform Power Loading Scenario
\thanks{This paper is part of the 03ED184 research project, implemented
within the framework of the "Reinforcement Programme of Human Research
Manpower" (PENED) and co-financed by National and Community Funds (20\% from
the Greek Ministry of Development-General Secretariat of Research and
Technology and 80\% from E.U.-European Social Fund)}
}


\author{Antonis~G.~Gotsis  \and
        Dimitris~I.~Komnakos \and
        Demosthenes D. Vouyioukas \and
        Philip~Constantinou}


\institute{A. Gotsis, D. Komnakos,  and Ph. Constantinou\at
              Mobile Radio-Communications Laboratory, ECE School, \\National Technical University of Athens, Greece. \\
              Tel.: +30-210-772 4196, Fax: +30-210-772 3851\\
              \email{\{gotsis,dkomna,fkonst\}@mobile.ntua.gr}           
\and D. Vouyioukas \at Computer and Communication Systems Laboratory\\ Dept. of
Information and Communication Systems Engineering, University of the Aegean\\
Karlovassi, Samos Island 83200, Greece\\ Tel: +30-22730-82270 Fax:
+30-22730-82269\\ \email{dvouyiou@aegean.gr}}

\date{Received: September, 23th, 2010/Revised: February, 3rd, 2011, March,20th 2012.}

\maketitle

\begin{abstract}
Adaptive Radio Resource Allocation is essential for guaranteeing high bandwidth
and power utilization as well as satisfying heterogeneous Quality-of-Service
requests regarding next generation broadband multicarrier wireless access
networks like LTE and Mobile WiMAX. A downlink OFDMA single-cell scenario is
considered where heterogeneous Constant-Bit-Rate and Best-Effort QoS profiles
coexist and the power is uniformly spread over the system bandwidth utilizing a
Uniform Power Loading (UPL) scenario. We express this particular QoS provision
scenario in mathematical terms, as a variation of the well-known generalized
assignment problem answered in the combinatorial optimization field. Based on
this concept, we propose two heuristic search algorithms for dynamically
allocating subchannels to the competing QoS classes and users which are
executed under polynomially-bounded cost. We also propose an Integer Linear
Programming model for optimally solving and acquiring a performance upper bound
for the same problem at reasonable yet high execution times. Through extensive
simulation results we show that the proposed algorithms exhibit high
close-to-optimal performance, thus comprising attractive candidates for
implementation in modern OFDMA-based systems.
\end{abstract}

\onehalfspacing

\section{Introduction} \label{sec:1}

Radio Resource Allocation (RRA) mechanisms are expected to play a key role in
emerging and future OFDMA-based multiuser wireless access networks. RRA aims at
simultaneously guaranteeing high utilization of the available system resources,
satisfying the individual Quality-of-Service (QoS) requirements of the
competing users, and maximizing overall system performance. In order to
accomplish these targets, an RRA or Frequency-Domain Packet Scheduling (under
the LTE terminology) technique exploits the differentiated
channel conditions experienced by the various users over the available
bandwidth. In particular, a complete allocation decision comprises the specific
set of OFDM subchannels assigned to each user as well as the transmission
format, namely the amount of power and the modulation mode, for each resource
block~\cite{Pi06}.

\subsection{Background and Related Work}

Adaptive RRA aims at positively exploiting the rate differentiation that occurs on two levels
closely related to respective physical phenomena: the single-user centric
variation of the achieved rate on a subchannel basis and the multi-user centric
differentiation of the rate achieved by each user on each subchannel. The
former phenomenon is related to the inherent frequency selectivity of the
wideband wireless medium, while the latter to the so-called multi-user
diversity caused by the statistical independence of the corresponding
subchannels. Therefore, by assigning to each user the subchannels
that experience favorable channel (and thus rate) conditions, we expect to significantly
improve system performance.

In \cite{ViTs02} the above concepts were thoroughly presented first for single- and then for multi-antenna wireless mobile systems. According to this concept each receiver monitors the experienced SNR levels, feeds them back to the BS, while the BS schedules transmissions and adapts users' bit rates depending on the particular channel quality reports. Similar arguments were raised in \cite{NaBa00} where it was demonstrated that for 2G/3G systems the cellular spectral efficiency may significantly improve, even double at certain conditions, when the BS utilizes the per-user channel/rate information. In \cite{BoGr07} the ideas were extended to multi-channel OFDM wireless systems like the one we examine in this paper. The widely used term "opportunistic" bears a strong relation with our work, since we tend to allocate subchannels with high channel conditions (near to their peak) to the corresponding users (frequency-domain opportunism), while in \cite{ViTs02} a similar policy is employed in the time-domain.

The works of \cite{WoCh99} and \cite{RhCi00} were the first to introduce an
optimization framework for handling multi-user OFDM resource allocation
problems, paving the way for an extensive utilization of concepts and methods
addressed in the engineering optimization field, towards efficiently assigning
system resources to active users. The original formulations of
the respective system power minimization~\cite{WoCh99} and sum-throughput
maximization~\cite{RhCi00} problems indicated the hard large-scale non-linear
integer nature of the underlying RRA decision problems, rendering the
straightforward discovery of the optimal allocation practically impossible.
Towards relaxing the complexity of the related problems, a plethora of
simplifying approaches has been proposed in the literature aiming at the
development of suboptimal, yet efficient and computationally tractable
allocation routines~\cite{KiLi03,KiLe06}. The works in \cite{ChHo09} and \cite{SaAn09} provide extensive surveys on the methodological and algorithmic aspects of the particular research area. A popular assumption which was first proposed in~\cite{JaLe03} and further elaborated in~\cite{ZhLe04} and
\cite{LeCh08} regarded the uniform power loading (UPL) of the system
subchannels. Under a known power distribution, the achieved bit-rates for all
the possible subchannel/user combinations can be precalculated, and thus the
original multidimensional RRA problems resort to exclusive subchannel
allocation problems. From a system perspective, OFDMA is considered the major transmission and multiple access technology for
modern wireless networks such as 3GPP-LTE~\cite{HoTo09}, Mobile
WiMAX~\cite{AnGh07} and also plays a key role in the new paradigm of Cognitive
Radio Networking~\cite{Ho09}. In particular, LTE-oriented works may be found in \cite{PoKo06,MoPe08},
WiMax-oriented ones in \cite{YaBe10,CaSa11,KhBo11} and cognitive networking related in \cite{ZhLe09,TaCh11} correspondingly.


\subsection{Contributions and Novelty}
In this paper we examine a single-cell downlink OFDMA system scenario supporting
multiple QoS profiles under the UPL assumption. Unlike single-profile studies,
which have attracted enormous interest in the related literature, the heterogeneous problem,
which is more interesting and realistic has not been given great attention. In \cite{HoTe98} a mixed CBR/VBR scenario was introduced for the single-user case, while in our work we consider multiple users demanding mixed services. The multi-user case was studied in \cite{WaHw04,YuZh07} where suboptimal allocation algorithms were proposed. However the UPL hypothesis have not been taken into account as in our work and the efficiency of the algorithms compared to the maximum achieved performance (that is, the optimal one) was not demonstrated. Finally, in \cite{ZhLe04} an efficient UPL-based resource allocation algorithm was proposed guaranteeing a set of minimum bit rates, while in our paper a more realistic QoS scenario is assumed, comprising mixed CBR and Best-Effort traffic profiles.

In particular, in this work we employ two optimization approaches to our problem. The former is based on Integer
Linear Programming (ILP) and allows us to find: (a) the exact optimal resource
allocation decision as well as the maximum sum-rate performance in reasonable,
yet high execution time, and (b) a performance upper bound in polynomial time
solving the LP-relaxed version of the IP formulation. The second approach is based on our observation that the specific heterogeneous
formulation bears similarities with a well-known combinatorial optimization
problem, the Generalized Assignment Problem (GAP) \cite{MaTo90}. In order to
efficiently solve the GAP formulation we devise two heuristic schemes, which
are executed with polynomially bounded computational cost. The development is
mainly inspired by the ideas utilized in the respective GAP heuristics found in
the related literature. The examined schemes are tested in an
extensive set of realistic computational-based network scenarios, in terms of
the achieved sum-rate performance. The simulation results show that the devised heuristic schemes: (i) outperform drastically a semi-random subchannel assignment approach, and (ii) perform close to the exact optimal allocation, while their complexity cost is significantly lower.

\subsection{Paper Structure}
The remaining of the paper is structured as follows. In Sec.~\ref{sec:2} we
present the adopted system model and the holding assumptions, state the
multiple-QoS profile RRA problem and formulate it as a Binary Integer Linear
Programming optimization problem. In Sec.~\ref{sec:3} we discuss how we can
express our problem as a variation of the GAP combinatorial format, and which
are the advantages of this approach, while in Sec.~\ref{sec:4} we devise two
heuristic allocation algorithms inspired by the GAP concept and discuss
computational complexity issues. In Sec.~\ref{sec:5} we present the results of
the computational experiments and comment on them, while in Sec.~\ref{sec:6} we
summarize our work and propose possible extensions of it.

\section{System Model and Mathematical Formulation}\label{sec:2}

\subsection{System Description and Assumptions}

The downlink (DL) of an OFDMA cell is considered in the context of this work,
where a Base Station (BS) located at the center of a 2D area is fully
responsible for allocating system subchannels (or equivalently physical
resource blocks according to the LTE terminology) to the competing users. The
available bandwidth comprises $N$ mutually-orthogonal subchannels (or resource
blocks) forming the set $\mathcal{S} = \left\{ {1,2, \ldots ,n, \ldots N}
\right\}$. The available transmission power $P_{bs}$ is uniformly spread over
the bandwidth, namely, each subchannel $n$ is loaded with an equal amount of
power given by ${P_n} = {{{P_{bs}}} \mathord{\left/
 {\vphantom {{{P_{bs}}} N}} \right.
 \kern-\nulldelimiterspace} N}$. We assume that $K$ users are present in the
 cell, forming the corresponding set $\mathcal{U} = \left\{ {1,2, \ldots ,k, \ldots K}
 \right\}$, and that each user is assigned one of the two available QoS profiles/classes,
 the Constant-Bit-Rate (CBR) or the Best-Effort (BE) one. Users belonging to
 the {CBR-class} subset (${{\cal U}_{{\rm{CBR}}}}$) demand a specific constant service rate
 denoted by $R_k^{\min }(\forall k \in {{\cal U}_{{\rm{CBR}}}})$, representing
 Voice/Video-like services. On the other hand, users belonging to the {BE-class} subset (${{\cal
 U}_{{\rm{BE}}}}$) have
 infinitely backlogged queues and no strict QoS guarantee, reflecting FTP-downloading
 services.

 In order to perform adaptive resource allocation in the above typical multi-user OFDMA system setup the following key operations/assumptions are supported:
 \begin{itemize}
   \item Each user estimates the BS-user link channel quality over the available subchannels and feeds back this information to the BS through an error-free and delay-less uplink signaling channel.
   \item Due to the wideband and multi-user nature of the transmission medium \cite{TsVi04} each user $k$ achieves different rate performance over each subchannel $n$, given by $r_{n,k}$ bits/symbol. Given that the BS exactly knows the wideband and multi-user channel response as well as the allocated power per subchannel $P_n$, the $r_{n,k}$ bit rates are pre-calculated (please refer to the discussion in Appedix~\ref{sec:app1}).
   \item Based on the above 2D rate matrix and the QoS targets, the BS decides the exact subchannel set assignments for each user. This information is also transmitted to each user in order for the useful data to be correctly decoded.
   \item Each subchannel should be allocated to a single user, thus avoiding intra-cell inter-user interference~\cite{Pi06}.
 \end{itemize}

%
%
%

\subsection{Mathematical Formulation and Optimal Solution}

We introduce $N\cdot K$ binary integer variables notated by $\rho_{n,k}$ where
$\rho_{n,k}=1$ if the $n^{th}$ arbitrary subchannel is allocated to the
$k^{th}$ arbitrary user, else equals zero. The data-rate $r_{n,k}$ supported on
each subchannel/user combination can be pre-calculated based on the abstraction
modeling function given in Appedix~\ref{sec:app1}. We are now able to formally define the RRA problem, which constitutes the
identification of the allocation decision that maximizes the cell sum-rate
($R_{tot}$), satisfies the individual data-rate constraints for the CBR-class
users and preserves subchannels orthogonality. Eq.\eqref{eq:ILPform_of} express
the system-wise objective function, Eq.\eqref{eq:ILPform_concbr} the $K_1
(\left| {{\mathcal{U}_{{\text{CBR}}}}} \right|)$ QoS constraints, and
Eq.\eqref{eq:ILPform_consys} the $N$ subchannel-sharing-avoidance system
constraints. Power availability and minimum BER constraints are handled
implicitly as explained in Appendix~\ref{sec:app1}. Although the generalized
bandwidth and power allocation problem is non-linear~\cite{WoCh99}, the UPL
hypothesis allows us to recast RRA in a linear form, as shown in
Eq.\eqref{eq:ILPform}. Actually, due to the exclusive assignment
of each subchannel to one user, the UPL variation of RRA is actually an Integer
Linear Programming (ILP) optimization problem. In fact we are dealing with an
ILP problem, including $N \cdot K$ binary variables and $\left|
{{\mathcal{U}_{{\text{CBR}}}}} \right| + N$ equality constraints, where $N\sim
100$ and $K\sim 10 -20$ under a typical network setup~\cite{HoTo09}. Despite
the fact that ILP problems are still hard to solve due to their NP-hard nature,
efficient methods and solvers like CPLEX, SCIP or GUROBI are available and may
be employed for problems of such scale.

Finally, an upper bound on the system performance could be estimated by
relaxing the integrality constraints or equivalently allowing for one or more
subchannels to be shared among users. For this case an LP solver could solve
the specific problem with 3rd order polynomial complexity cost. Note however
that the LP relaxation approach is able to provide us only with the system
performance (in fact the theoretical achieved upper bound) and not with an
implementable allocation decision due to the violation of the
subchannel-sharing constraints set.  We have to emphasize the implementation
limitations of both IP and LP approaches, underlying that their value is mostly
theoretical. For this reason we proceed with developing simpler allocation
schemes in the following two sections.

\begin{subequations}
\begin{gather}
  \mathop {\max }\limits_\text{\boldmath{$\rho$}}  {R_{tot}} = \sum\limits_{n \in \mathcal{S}} {\sum\limits_{k \in \mathcal{U}} {{r_{n,k}} \cdot {\rho _{n,k}}} }  = \sum\limits_{k \in {\mathcal{U}_{{\text{CBR}}}}} {R_{\min }^k}  + \sum\limits_{n \in \mathcal{S}} {\sum\limits_{k \in {\mathcal{U}_{{\text{BE}}}}} {{r_{n,k}} \cdot {\rho _{n,k}}} }  \hfill \label{eq:ILPform_of}\\
  {\text{subject to}} \hfill \nonumber\\
  \sum\limits_{n \in \mathcal{S}} {{r_{n,k}} \cdot {\rho _{n,k}}}  = R_{\min }^k,\forall k \in {\mathcal{U}_{{\text{CBR}}}} \hfill \label{eq:ILPform_concbr}\\
  \sum\limits_{n \in \mathcal{S}} {{\rho _{n,k}}}  = 1,\forall n \in \mathcal{S},\text{\boldmath{$\rho$}} \in {\mathbb{I}^{N \times K}},\mathbb{I} = \left\{ {0,1} \right\}
  \hfill \label{eq:ILPform_consys}
\end{gather}
\label{eq:ILPform}
\end{subequations}

\section{Radio Resource Allocation as a Variation of the Generalized Assignment Problem}\label{sec:3}

The Generalized Assignment Problem (GAP) along with its variations is a
well-studied special formulation of Combinatorial Optimization. Although GAP is
still an NP-hard problem, several efficient approximate heuristic algorithms
have been proposed for solving it~\cite{MaTo90,MoRo04} due to its wide
application in real-world problems. A Knapsack-based definition of GAP proposed
in~\cite{MaTo90} constitutes the assignment of items to knapsacks in order to
maximize the overall profit while not exceeding the capacity constraints of
each knapsack. The assignment of each item to a knapsack incurs different
profit and induces different cost. Note that an item can be obviously assigned
to a single knapsack. A slightly different definition is given in
\cite{MoRo04}, where one searches for the best scheduling of tasks or jobs to
agents, in order to minimize the overall processing cost and do not violate the
available resource budget of each agent. Again a task cannot be split into
multiple agents. Finally, a variation of the GAP, known as the Covering
Assignment Problem (CAP) is given in~\cite{FoWi97} which applies to a dairy
industry distribution scenario. Specifically, CAP regards the optimal
distribution procedure of the milk produced by several farms to a set of
factories, in order to minimize the cost of processing and simultaneously
satisfy the minimum resource demand of each factory.

Towards expressing our RRA problem as a GAP, we first define the correspondence
of an item/task/farm to a subchannel and a knapsack/agent/factory to a user of
the system. RRA may then be seen as the optimal exclusive allocation of each
subchannel to a single user in order to maximize the sum-rate (profit) function
and provide minimum rate (resource) assignments for a user subset.We also have
to mention that our formulation possesses several distinctive features compared
to the aforementioned classical GAP/CAP approaches:
\begin{itemize}
  \item The constraint expressions are tight whereas in GAP and CAP they correspond
  to left-hand-side or right-hand-side inequalities
  \item The constraint expressions are imposed on a subset of users and not
  over the complete users' set
  \item the profit/weight/cost factors are equal ($r_{n,k}$) and depend on both
  subchannel and user indexes through the experienced rate/channel conditions.
\end{itemize}
As far as GAP and its variations, several efficient heuristic algorithms could
be found in the literature. In~\cite{MaTo90} the authors proposed a scheme
which was based on the construction of an initial effective feasible solution
and its subsequent improvement through item reallocations. In~\cite{Wi97} the
author proposed a ``dual" algorithm, according to which the optimal solution
was approached from the exterior of the feasible region. Finally,
in~\cite{FoWi97} the authors devised similar schemes for the CAP variation. In
the following section we devise two algorithms, inspired by the previous
GAP/CAP works. We remark that due to the differences between our RRA problem
and the classical formulations, none of the existing algorithms could be
employed as-is. Moreover, although other solution approaches may be found in
the literature, such as the Lagrangean Relaxation, Branch-and-Bound,
Metaheuristics, etc. (see for example a survey in~\cite{CaWa92}) we focus on
the heuristic solution approach since it is appealing from an implementation
point of view. One should bear in mind that in a practical network scenario the
complete resource assignment must be updated every transmission frame. For
example, regarding contemporary wireless access systems like LTE and WiMAX, the
update rate should be of the order of 0.5--1 msec~\cite{HoTo09}.

\section{Heuristics for Solving the Resource Allocation Problem}\label{sec:4}

\subsection{Heuristic I: Approaching the Optimal Allocation from the Interior of the Feasible
Space} \label{sec:4a}

\paragraph{Description} The first heuristic is mainly inspired by the Knapsack
solution approach given in~\cite{MaTo90}. Resource Allocation is performed in 2
phases, that is the construction of an initial feasible solution and the search
for a gradually improved allocation. Hereafter, when we refer to a ``feasible"
allocation decision, we will mean that the $\left|
{{\mathcal{U}_{{\text{CBR}}}}} \right|$ data rate constraints are satisfied. As
far as the remaining $N$ orthogonality constraints, these will be implicitly
met, since at each subchannel allocation step, the subchannel which is assigned
to a user will be removed from the available bandwidth ``pool". Hence, we
should not confuse the use of the feasibility term with the one regarding the
LP relaxation given in Sec.\ref{sec:2}. The complete algorithm including all
the intermediate steps for each phase is given after the description of its key
features.

During the \text{1st phase}, our primary objective is to construct a feasible
RRA solution, that is, satisfy the strict data-rate constraints of all CBR
users, while secondarily the respective allocation has to be as efficient as
possible in terms of the overall system performance. Towards the 1st point we
prioritize the CBR-class subchannel allocation procedure over BE-class, thus
decoupling the inter-class problem to two consecutive intra-class sub-problems.
Concerning the secondary target, we state that if the above QoS constraints are
satisfied by utilizing the minimum amount of the available Tx power as well as
the ``best" subchannels in terms of the achieved channel/rate conditions, then
plenty of resources will be available for BE-class users. Such a policy would
enhance system performance, since BE users are the ones that contribute to the
objective function, due to the predetermined static nature of the CBR-class
users' data-rates. Hence, after the necessary initializations and declarations
(\textbf{Step 1}), an iterative joint subchannel/user allocation is applied
over the CBR users subset ignoring BE users (\textbf{Step 2}). The selection
criterion for picking the best CBR subchannel-user pair at each iteration is
dual: on the one hand the user experiencing the minimum averaged achieved
data-rate over the remaining available subchannels set is prioritized and on
the other hand the ``best" subchannel in terms of rate performance is promoted.
The user-selection criterion prioritizes the users with the worse rate
conditions, which are expected to consume the largest portion of resources, and
by allocating to them their most efficient subchannels we succeed into
constraining the overall amount of CBR-class resources consumption.

Subsequently, we finalize the initial feasible RRA solution by assigning the
unallocated resources to the BE users (\textbf{Step 3}). Since no QoS
requirements are imposed for the particular class, the simple policy of
best-rate user allocation is also the optimal one (see also~\cite{JaLe03}). Up
to now, we have provided a feasible allocation decision, however we argue that
we could further improve it since the previous steps may induce suboptimality
to the sum-rate performance. Note that this is caused by the ``greedy" (local)
nature of the heuristic allocations during Step 2. Towards the above purpose we
employ a series of subchannels swaps (\textbf{Step 4}). Specifically, we
perform a subchannel interchange between 2 users if and only if the system
performance is enhanced and simultaneously the feasibility for any CBR user is
not violated. Due to the presence of multiple QoS classes, we have to
discriminate between $2^2=4$ possible swap scenarios (2 classes are considered
in this work). Finally in order to limit the number of interchanges (and thus
the execution complexity) we perform only one round of comparisons and possible
reallocations by a single sweep of the set of system users. We remark that by increasing the comparison rounds, a marginal additional performance gain is expected, since already at each user iteration the owner's subchannels are compared with the subchannels of all the other users.

The last subprocedure given in \textbf{Step 5} comprises the release of
possible redundant subchannels of CBR-class users due to the active nature of
the related constraints and the reallocation of them to a BE user. We put an
emphasis on the fact that for CBR users surplus allocated data-rates are
ignored.

\paragraph{Complete Algorithm}

\paragraph{\textbf{Step 1}}  \emph{Initializations-Declarations}\\[3pt]
Let ${\left\{ {{r_{n,k}}} \right\}_{n \in \mathcal{S},k \in \mathcal{U}}}$ the
known achieved data-rates for all subchannel/user combinations and ${\left\{
{{\rho _{n,k}}} \right\}_{n \in \mathcal{S},k \in \mathcal{U}}}$ the set of the
allocation variables.\\[2pt]
Define: $lh{s_k} = \sum\limits_{n = 1}^N {{r_{n,k}} \cdot {\rho _{n,k}}} \text{
and } {b_k} = R_{\min }^k, \forall k \in {\mathcal{U}_{{\text{CBR}}}}$.

\paragraph{\textbf{Step 2}}  \emph{Subchannel Assignment to the CBR-class users -- Feasible Allocation}\\[3pt]
Define the available subchannels pool $\mathcal{N}$ and the subset of the
unsatisfied CBR users $\mathcal{K}$ as:\\[2pt]
$\mathcal{N} = \left\{ {n \in \mathcal{S}:\sum\limits_{k = 1}^K {{\rho _{n,k}}}
= 0} \right\},\mathcal{K} = \left\{ {k \in {\mathcal{U}_{{\text{CBR}}}}:lh{s_k}
- {b_k} < 0} \right\}$.\\[2pt]
Update $lh{s_k} = \sum\limits_{n = 1}^N {{r_{n,k}} \cdot {\rho _{n,k}}} \text{
and } {b_k} = R_{\min }^k, \forall k \in {\mathcal{U}_{{\text{CBR}}}}$.\\[2pt]
Pick the subchannel/user combination according to:\\
${k^*} = \arg \mathop {\min }\limits_{k \in \mathcal{K}} \left\{ {\left( {{1
\mathord{\left/
 {\vphantom {1 {\left| \mathcal{N} \right|}}} \right.
 \kern-\nulldelimiterspace} {\left| \mathcal{N} \right|}}} \right) \cdot \sum\limits_{n \in \mathcal{N}} {{r_{n,k}}} } \right\},{n^*} = \arg \mathop {\max }\limits_{n \in \mathcal{N}} \left\{ {{r_{n,{k^*}}}}
 \right\}$ and\\[2pt]
Perform the subchannel allocation: ${\rho _{{n^*},{k^*}}} = 1$\\[2pt]
If $\mathcal{K} \ne \emptyset$ Repeat Step 2 else Break the Loop and Go To Step
3

\paragraph{\textbf{Step 3}}\emph{Subchannel Allocation to BE-class users -- Finalization of the Feasible Solution}\\[3pt]
Assign each unallocated subchannel to the ``best" users in terms of the
achieved rate: ${\text{For each }}n \in \mathcal{N},{\text{ set }}{\rho
_{n,{k^*}}} = 1{\text{, where }}{k^*} = \arg \mathop {\max }\limits_{k \in
{\mathcal{U}_{{\text{BE}}}}} \left\{ {{r_{n,k}}} \right\}$\\[2pt]

\paragraph{\textbf{Step 4}}\emph{Subchannel Swapping -- Improved Solution}\\[3pt]
For every user $k \in \mathcal{U}$ (either {CBR} or {BE}) repeat the following
procedure serially:\\[2pt]
(a) Find his/her allocated subchannels set:
${\mathcal{I}_k} = \left\{ {n \in \mathcal{S}:{\rho _{n,k}} = 1} \right\}$\\[2pt]
(b) Find the complementary users/subchan. sets : $\mathcal{U}'{_k}  =
\mathcal{U}\backslash \left\{ k \right\},{{\mathcal{I}'}_k} =
\mathcal{S}\backslash {\mathcal{I}_k}$\\[2pt]
(c) For all possible combinations $\left\langle {\left( {n,k}
\right){\text{,}}\left( {n',k'} \right)} \right\rangle$, where $n \in
{\mathcal{I}_k},n' \in {{\mathcal{I}'}_k},k \in \mathcal{U}'{_k}$ check the
inner conditional expression depending on the active scenario:\\[2pt]
(c.1) If ${k \in {\mathcal{U}_{{\text{CBR}}}}{\text{ and  }}k' \in
{\mathcal{U}_{{\text{CBR}}}}}$: ${r_{n',k}} > {r_{n,k}}{\text{ \&\&
}}lh{s_{k'}} + {r_{n,k'}} - {r_{n',k'}} \geqslant {b_{k'}}{\text{ }}$ \\ \text{
} \hspace{125pt} or ${r_{n,k'}}
> {r_{n',k'}}{\text{ \&\& }}lh{s_k} + {r_{n',k}} - {r_{n,k}} \geqslant
{b_k}$\\[2pt]
(c.2) If ${k \in {\mathcal{U}_{{\text{BE}}}}{\text{ and  }}k' \in
{\mathcal{U}_{{\text{BE}}}}}$: \hspace{8pt} ${r_{n',k}} - {r_{n,k}} +
{r_{n,k'}} - {r_{n',k'}}
> 0$\\[2pt]
(c.3) If $k \in {\mathcal{U}_{{\text{CBR}}}}{\text{ and }}k' \in
{\mathcal{U}_{{\text{BE}}}}$: \hspace{3pt} ${r_{n,k'}} > {r_{n',k'}}{\text{
\&\&
}}lh{s_k} + {r_{n',k}} - {r_{n,k}} \geqslant {b_k}$\\[2pt]
(c.4) If ${k \in {\mathcal{U}_{{\text{BE}}}}{\text{ and  }}k' \in
{\mathcal{U}_{{\text{CBR}}}}}$: \hspace{0.5pt} ${r_{n',k}} > {r_{n,k}}{\text{
\&\& }}lh{s_{k'}} + {r_{n,k'}} - {r_{n',k'}} \geqslant {b_{k'}}$\\[2pt]
If any of the conditions is TRUE employ the corresponding resource swapping:\\[2pt]
${\rho _{n,k}} = 0,{\rho _{n',k'}} = 0,{\rho _{n',k}} = 1,{\rho _{n,k'}} = 1$.

\paragraph{\textbf{Step 5}}\emph{Release of CBR-class Subchannels for BE-class}\\[3pt]
For every CBR user $k \in {\mathcal{U}_{{\text{CBR}}}}$:\\[2pt]
(a) Find his/her allocated subchannels set:
${\mathcal{I}_k} = \left\{ {n \in \mathcal{S}:{\rho _{n,k}} = 1} \right\}$\\[2pt]
(b) For each $n \in {\mathcal{I}_k}{\text{ if }}lh{s_k} - {r_{n,k}} \geqslant
{b_k}$ release it from this user and allocated it to the best BE-user: ${\rho
_{n,k}} = 0,{\rho _{n,{k^*}}} = 1,{\text{ where }}{k^*} = \arg \mathop {\max
}\limits_{k \in {\mathcal{U}_{{\text{BE}}}}} \left\{ {{r_{n,k}}} \right\}$

\subsection{Heuristic II: A Dual Approach of the Optimal Allocation}  \label{sec:4b}

\paragraph{Description} An alternative heuristic algorithm inspired by the works of \cite{Wi97} and
\cite{FoWi97}, where the optimal allocation decision is approached from the
exterior of the feasible space, is proposed in the present subsection. The
``dual" approach shares also several similarities with the Lagrangean
Relaxation technique, which dualizes the hard constraints. Specifically, the
solution is approached on 2 phases: the construction of the optimal
unconstrained subchannel allocation and the feasibility transformation of the
initial decision through a series of subchannel reallocations. All the
intermediate allocations generated by the dual approach are infeasible except
for the last one. Note, that a similar algorithm for a single-QoS profile RRA
problem has been proposed in~\cite{ZhLe04}. We now provide a summary of the
algorithm as well as a complete step-by-step description of the dual approach
in the context of our RRA problem.

During the \text{1st phase}, the necessary initializations and declarations
(\textbf{Step 1}) as well as the optimal unconstrained subchannel allocation
are performed (\textbf{Step~2}). The optimal solution to the corresponding
unconstrained problem is extracted by simply allocating each subchannel to the
user that experiences the maximum data-rate on it. Obviously, the sum-rate
performance of the best-user allocation policy is an upper bound on any
QoS-constrained scenario. The \text{2nd phase} on the other hand comprises a
series of subchannel reallocations aiming at satisfying the minimum QoS targets
or equivalently rendering the solution feasible (\textbf{Step
3})\footnote{During the reallocation subprocedure followed here, subchannels
are removed from one user and given to another, whereas in the 1st heuristic
subchannels interchanges occur}. The selection of user/subchannel pairs
participating in each reallocation cycle is dictated by the efficiency metric
shown in Eq.\eqref{eq:Realloc_crit}, where $n$ stands for a subchannel
candidate for reassignment, $k^*(n)$ for the current owner of the subchannel
(as determined by Step 2), and $k$ the index of the candidate reassigned user.
\begin{equation}\label{eq:Realloc_crit}
{a_{n,k}} = {\left. {\dfrac{{{r_{n,{k^*}\left( n \right)}} - {r_{n,k}}}}
{{{r_{n,k}}}}} \right|_{n \in \mathcal{N},k \in \mathcal{J}'}}
\end{equation}
The nominator of the metric expresses the decrease in sum-rate performance due
to the reallocation, whereas the denominator the increase of the the infeasible
user data-rate or equivalently the decrease of the distance from feasibility.
Thus, by minimizing this metric we simultaneously harm as less as possible the
original objective function value and approach as fast as possible the
feasibility region. Notice, however that a specific resource-block reallocation
is possible if the minimum QoS target for the owner-user is not violated. The
particular iterative procedure ends when all CBR users' demands are satisfied.
Finally, the last step of Phase 2 (\textbf{Step 4}) constitutes the release of
redundant subchannels (if any) from the CBR-class users, similarly to the
procedure followed in the 1st Heuristic.

\paragraph{Complete Algorithm}

\paragraph{\textbf{Step 1}}  \emph{Initializations -- Declarations}\\[3pt]
Let ${\left\{ {{r_{n,k}}} \right\}_{n \in \mathcal{S},k \in
\mathcal{U}}},{\left\{ {{\rho _{n,k}}} \right\}_{n \in \mathcal{S},k \in
\mathcal{U}}},{\left\{ {lh{s_k}} \right\}_{k \in
{\mathcal{U}_{{\text{CBR}}}}}},{\left\{ {{b_k}} \right\}_{k \in
{\mathcal{U}_{{\text{CBR}}}}}}$ as in Heuristic I.\\[2pt]

\paragraph{\textbf{Step 2}}  \emph{Optimum Unconstrained (QoS-unaware) Allocation}\\[3pt]
Best-Rate Allocation: For each $n \in \mathcal{S},{\text{ set }}{\rho
_{n,{k^*}}} = 1,{\text{ where }}{k^*} = \arg \mathop {\max }\limits_{k \in
\mathcal{U}} \left\{ {{r_{n,k}}} \right\}$\\[2pt]
Minimum QoS satisfaction check: If for all $k \in
{\mathcal{U}_{{\text{CBR}}}}{\text{ : }}lh{s_k} \geqslant {b_k}$ Then Go To
Step 4 Else Go To Step 3.

\paragraph{\textbf{Step 3}}  \emph{Subchannels Reallocation -- QoS Satisfaction}\\[3pt]
Define the following subchannels and users subsets:\\[2pt]
${\mathcal{J}_{{\text{CBR}}}} = \left\{ {k \in
{\mathcal{U}_{{\text{CBR}}}}:lh{s_k} \geqslant {b_k}} \right\} \subseteq
{\mathcal{U}_{{\text{CBR}}}}$: the subset of CBR users for which minimum QoS is
met.\\[2pt]
${{\mathcal{J}'}_{{\text{CBR}}}} = {\mathcal{U}_{{\text{CBR}}}}\backslash
{\mathcal{J}_{{\text{CBR}}}}$: the complementary subset of undersatisfied CBR
users.\\[2pt]
${{\mathcal{J}}_{{\text{BE}}}} = \left\{ {k \in
{\mathcal{U}_{{\text{BE}}}}:lh{s_k} > 0} \right\} \subseteq
{\mathcal{U}_{{\text{BE}}}}$: the subset of BE users allocated at least one
subchannel or equivalently posses non-zero data rate.\\[2pt]
$\mathcal{J} = {{\mathcal{J}}_{{\text{CBR}}}} \cup
{{\mathcal{J}}_{{\text{BE}}}}{\text{ }}$: the subset of users from which we can
remove subchannels\\[2pt]
$\mathcal{J}' = {{\mathcal{J}'}_{{\text{CBR}}}}$: the subset of users to which
we must add subchannels in order to meet the minimum QoS requirements.
$\mathcal{N} = \left\{ {n \in \mathcal{S},k \in \mathcal{J}:{\rho _{n,k}} = 1}
\right\}$: the subchannels pool from which we can extract subchannels and
reallocate to the users subset $\mathcal{J}'$.\\[4pt]
Repeat the following procedure while there exist unsatisfied CBR users, namely
as long as $\mathcal{J}' \ne \emptyset$:\\[2pt]
(a) For each subchannel $n \in \mathcal{N}$ identify the owner ${k^*}\left( n
\right) = \left\{ {k \in \mathcal{J}:{\rho _{n,k}} = 1} \right\}$.\\[2pt]
(b) Compose the 2D Reallocations Array ${{\mathbf{A}}_{\left| \mathcal{N}
\right| \times \left| {\mathcal{J}'} \right|}} = \left[ {{a_{n,k}}} \right]$,
where:\\[2pt]
${a_{n,k}} = \left\{ \begin{gathered}
  \dfrac{{{r_{n,{k^*}\left( n \right)}} - {r_{n,k}}}}
{{{r_{n,k}}}},{\text{ if }}lh{s_{{k^*}\left( n \right)}} - {r_{n,{k^*}\left( n \right)}} \geqslant {b_{{k^*}\left( n \right)}}{\text{ and }}{r_{n,k}} > 0 \hfill \\
   + \infty \hspace{45pt} {\text{,  elsewhere}} \hfill \\
\end{gathered}  \right.$\\[2pt]
(c) Check all the elements of the array:\\[2pt]
If for all elements ${a_{n,k}} =  + \infty$, Break the Loop and Go To Step
4.\\[2pt]
Else pick the subch./user combination according to $\left\{ {n',k'} \right\} =
\arg \mathop {\min }\limits_{n \in \left| \mathcal{N} \right|,k \in
\mathcal{J}'} {\mathbf{A}}$ and perform the reallocation: ${\rho
_{n',{k^*}\left( {n'} \right)}} = 0,{\rho _{n',k'}} = 1$. \\[3pt]
(d) Update the related quantities/subsets:
$\mathcal{J},\mathcal{J}',\mathcal{N},lh{s_k},k \in
{\mathcal{U}_{{\text{CBR}}}}$.

\paragraph{\textbf{Step 4}}  \emph{Redundant CBR-class Subchannels Reallocation}\\[3pt]
Define the following subchannels and users subsets:\\[2pt]
$\mathcal{J} = \left\{ {k \in {\mathcal{U}_{{\text{CBR}}}}:lh{s_k} > {b_k}}
\right\}$: the subset of CBR users assigned redundant data-rate (over-satisfied
users).\\[2pt]
$\mathcal{N} = \left\{ {n \in \mathcal{S},k \in \mathcal{J}:{\rho _{n,k}} = 1}
\right\}$: possible redundant subchannels pool.\\[2pt]
Repeat the following procedure while there exist over-satisfied users, namely
as long as $\mathcal{J} \ne \emptyset$:\\[2pt]
For each $n \in \mathcal{N}$:\\[2pt]
(a) find the owner ${k^*}\left( n \right) = \left\{ {k \in \mathcal{J}:{\rho
_{n,k}} = 1} \right\}$\\[2pt]
(b) If $lh{s_{{k^*}}} - {r_{n,{k^*}}} \geqslant {b_{{k^*}}}$ remove the
subchannel $n$ from the owner ${k^*}\left( n \right)$ and reallocate it to the
``best" BE user in terms of achieved data-rate $k'$:\\[2pt]
${\rho _{n',{k^*}\left( {n'} \right)}} = 0,{\rho _{n',k'}} = 1$, where
$k'\left( n \right) = \arg \mathop {\max }\limits_{k \in
{\mathcal{U}_{{\text{BE}}}}} \left\{ {{r_{n,k}}} \right\}$, Else Go To the next
subchannel.

\subsection{Computational Complexity Estimation}
An estimation of the computational complexity of the proposed optimal and
suboptimal schemes follows. With respect to the heuristic algorithms we take
into account the involved searching, sorting and comparison operations. We
assume that the order of CBR and BE users is the same that is, $\left|
{{\mathcal{U}_{{\text{CBR}}}}} \right| \sim \left|{{\mathcal{U}_{{\text{BE}}}}}
\right| \sim K$ and that the number of subchannels is an order greater than the
number of active users, namely $N \gg K$.

\paragraph{Exhaustive and ILP/LP models} We first consider an exhaustive
search approach, where all the possible combinations of subchannels-to-users
assignments are examined. The particular procedure has an exponential
complexity order of $O(K^N)$, which is obviously computationally intractable.
As far as the exact optimal ILP model there is no guarantee for the execution
complexity, due to its NP-hard nature, however it is expected to be
significantly lower than the complete enumeration. The relaxed LP model on the
other hand has a provable third-order polynomial solution complexity with
respect to the number of the involved variables. Hence, in order to find the
performance upper bound of the RRA problem we have to spend a set of
computational operations of order $O\left( {{N^3}{K^3}} \right)$.

\paragraph{Heuristic I} Since the introductory Step 1 induces no complexity, we proceed directly with the next step.
We first employ a necessary pre-sorting operation regarding the achieved
data-rate values over all users, which costs $KN\log N$ operations (this is an
implementation issue, and this is why it was omitted from the previous
algorithmic description). Regarding Step 2 we perform at most $N$ iterations
for satisfying the CBR-class QoS constraints, and at each iteration we have to
select the worst user in terms of the average achieved data-rate with
complexity cost $K$ and locate its best subchannel with cost $logN$ (due to
pre-sorting), leading to a total cost of $NK + N\log N$ operations. Step 3
involves $N\log K$ operations for identifying the best-rate user for each
remaining subchannel. Step 4 is the most complex one, since it entails a series
of comparisons: without loss of accuracy we assume that each user is
preallocated on average ${N \mathord{\left/{\vphantom {N K}}
\right.\kern-\nulldelimiterspace} K}$ subchannels, and then for each user the
assigned ${N \mathord{\left/
 {\vphantom {N K}} \right.
 \kern-\nulldelimiterspace} K}$ subchannels are compared with the ${N \mathord{\left/
 {\vphantom {N K}} \right.
 \kern-\nulldelimiterspace} K}$ subchannels of the complementary $\left( {K - 1} \right)$
 users. All users are scanned for a full cycle, hence the whole process costs
 $\dfrac{N} {K} \cdot \dfrac{N} {K} \cdot \left( {K - 1} \right) \cdot K \approx
{N^2}$ operations. Finally, Step 5 involves at most $N$ subchannel releases,
and for each one the best-rate BE user must be located with cost $\log K$,
hence $N\log K$ operations are needed. Combining all the above algorithmic
steps we get $KN\log N + NK + N\log N + {N^2} + N\log K$ operations and after
performing several manipulations and approximations we resort to an estimated
computational complexity order of $O\left( {{N^2} + NK\log N} \right)$.

\paragraph{Heuristic II} Similarly to the 1st heuristic we first employ a pre-sorting operation for the
2D rates array which costs $KN\log N$ operations. Step 2 needs $N\log K$
operations in order to find the optimal unconstrained allocation. Step 3
contributes significantly to the overall complexity: at most $N$ subchannel
reallocations are performed and for each one a reallocation cost matrix
containing $NK$ elements must be devised and then searched for the combination
experiencing the minimal cost. Thus $N \cdot NK = {N^2}K$ operations are needed
to fulfil the particular step. Step 4 induces an additional cost of $N \log K$
operations as in the 1st heuristic. Accounting for all the algorithmic steps we
result to an overall approximate complexity order of $O\left( {{N^2} + NK\log
N} \right)$.

\section{Simulation Results and Discussion}\label{sec:5}

\paragraph{Simulation Setup}

The DL of a single-cell OFDMA-based packet data network is modeled and
simulated in the context of this work, while the selection of system parameters
reflects an LTE scenario~\cite{HoTo09}. A system bandwidth of 20~MHz is
assumed, consisting of $N$~=~100 orthogonal data subchannels. DL transmissions
occur on frame bursts of $T_f$~=~1~msec. A realistic pedestrian NLOS macro-cell
urban channel from the WinnerII models family is adopted~\cite{WINII07a}. The
PHY-abstraction function of Cioffi is utilized for associating each
channel-to-noise ratio level with an achieved data rate (see
Appendix~\ref{sec:app1}). Channel conditions are assumed perfectly known at the
BS, allowing for an opportunistic subchannel allocation. At the receiver side
transmissions are harmed due to thermal noise with power density of
$N_0$~=~$-174$ dBm/Hz.

As far as the traffic/QoS models, a CBR/BE dual-class scenario is formed. Each
CBR user requires $R_k^{\min}$= 36 {bits/OFDMA symbol}\footnote{For simplicity
reasons we may assume that an OFDMA frame carries one data-symbol. Thus all the
data-rate/throughput quantities may be identically expressed in bits/OFDMA
symbol/frame or simply in bits.}; the overall required CBR load is determined
by varying the number of users, where $\left| {{\mathcal{U}_{{\text{CBR}}}}}
\right|$~=~6--12. The number of BE users is held constant at $\left|
{{\mathcal{U}_{{\text{BE}}}}} \right|$~=~5. System capacity may be calculated
as $C_{max}=N \cdot c_{max}$~=~600~bits, where $c_{max}$ is the maximum
supported data-rate per subchannel, however the true achieved cell-rate is
expected to be lower due to the hard CBR rate constraints and BE-class users'
power shortage. The simulation parameters are summarized in
Table~\ref{ta:SimParam}.

%
%

\begin{table}[t]
\footnotesize  \caption{Simulation Parameters} \label{ta:SimParam} \centering
\vspace{6pt}
\begin{tabular}{lll}
 \hline\noalign{\smallskip}
\textbf{Quantity} & \textbf{Symbol} & \textbf{Value/Comment}  \\
\noalign{\smallskip}\hline\noalign{\smallskip}
Carrier Frequency & $f_c$ & 2.5 GHz \\
System Bandwidth & BW &  20 MHz \\
Subchannel Bandwidth & $\Delta f$ & 200 kHz \\
Number of Subchannels & $N$ & 100 \\
Transmission Error Rate & $P_e$ & $10^{-6}$ \\
Noise Power Density & $N_0$ & -174 dBm/Hz \\
Higher Tx Order per suchannel & $c_{max}$ & 6 bits/symbol \\
Cell Radius & $R_{cell}$ & 2 km \\
Channel Model & -- & WinnerII C3 BU-NLOS Macro\\
Users Distribution & -- & Uniform\\
Users Mobility Model & -- & Pedestrian (4 km/hr)\\
Number of OFDMA frames per drop & -- & 100\\
CBR QoS model & ${\left\{ {R_k^{\min }} \right\}_{k \in
{\mathcal{U}_{{\text{CBR}}}}}}$ & 36 bits/OFDMA symbol \\
CBR users & $K_1=\left| {{\mathcal{U}_{{\text{CBR}}}}} \right|$ & 6--12 \\
BE users & $K_2=\left| {{\mathcal{U}_{{\text{BE}}}}} \right|$ & 5 \\
Max Number of drops & $N_{drops}^{ulim}$ & 1000 \\
Min Number of drops & $N_{drops}^{llim}$ & 25 \\
Statistics Convergence Threshold & $\sigma_{norm}$ & 0.02 \\
\noalign{\smallskip}\hline
\end{tabular}
\end{table}

We consider five schemes which are compared below in terms of the achieved
sum-rate performance while we guarantee the CBR-class rates:
\begin{itemize}
  \item The exact optimal scheme extracted by solving the ILP optimization problem
  given in in Eq.~\eqref{eq:ILPform} (\textbf{IP})
  \item A scheme that provides us with a performance upper bound, obtained by
  solving the continuous relaxed version of the previous scheme (\textbf{LP})
  \item The 1st heuristic proposed in Sec.\ref{sec:4a} (\textbf{HEUR1}), for which we also run a version without using the swapping sub-procedure (\text{HEUR1 no swap})
  \item The 2nd heuristic proposed in Sec.\ref{sec:4b} (\textbf{HEUR2})
  \item A semi-random allocation algorithm  (\textbf{RANDOM}) which: (a) satisfies the target rates for the CBR-class users, by picking them one by one and assigning their best available channel until all requirements are met and, (b) assigns the remaining subchannels to the BE users randomly
\end{itemize}

All the algorithms are implemented in MATLAB and for the ILP/LP problems we
utilize the CPLEX solver~\cite{CPLEX}, calling it through the TOMLAB
interface~\cite{Tomlab99}. In order to capture the effect of different system
parameters to the achieved performance, we simulate 20 realistic system
scenarios by varying:
\begin{itemize}
   \item [(i)] the CBR-loading/QoS levels in the service area, by
   considering a different number of active CBR users ($K_1 = \left| {{\mathcal{U}_{{\text{CBR}}}}}
   \right|$~=~6,8,10,12)
   \item [(ii)] the average experienced SNR conditions in the cell, by
   tuning the ratio of the BS transmission power ($P_{bs}$) to the minimum required Tx power for guaranteeing feasibility regarding the $K_1$
   data-rate constraints ($P_{CBR}^{feas}(K_1)$) in each scenario ($P_{bs}/P_{CBR}^{feas}$~=~$2.0,2.5,\ldots,4.0$
   \footnote{Note that lower values for the power availability metric (e.g. 1.0 or 1.5) are not examined. This is justified
   by the fact that at lower SNR conditions, one or more CBR rate constraints can not be
   satisfied (``outage" conditions), rendering the QoS constrained problem infeasible. In order to cope with such situations, an adaptive power
   allocation strategy must be employed, like the one we proposed in our recently published work~\cite{GoKo10}.
   })
 \end{itemize}

We define an arbitrary realization or ``drop" as a system setup consisting of a
set of users randomly placed in the cell, for which their large-scale channel
conditions are held constant, whilst the small-scale conditions vary in time.
Each drop consists of 100 consecutive OFDMA frames, spanning $100 \cdot
T_f$~=~100~msec. For each scenario we simulate multiple statistically
independent drops and record the average sum-rate performance of each scheme.
The simulation is terminated if the normalized (to the mean) variance of the
performance statistics drops below the convergence target threshold (which is
set to 0.02) or if the maximum number of simulated drops (which is set to 1000)
has been reached~\cite{JeBa00}. Moreover, a minimum number of 25 drops are
executed in any scenario, in order to avoid transient effects. Thus, at least
$25 \cdot 100$ = 2,500 and at most $1000 \cdot 100$ = 100,000 optimization
problems are formulated and solved for each scenario.

\begin{figure}
\centering
\includegraphics[width=0.75\textwidth]{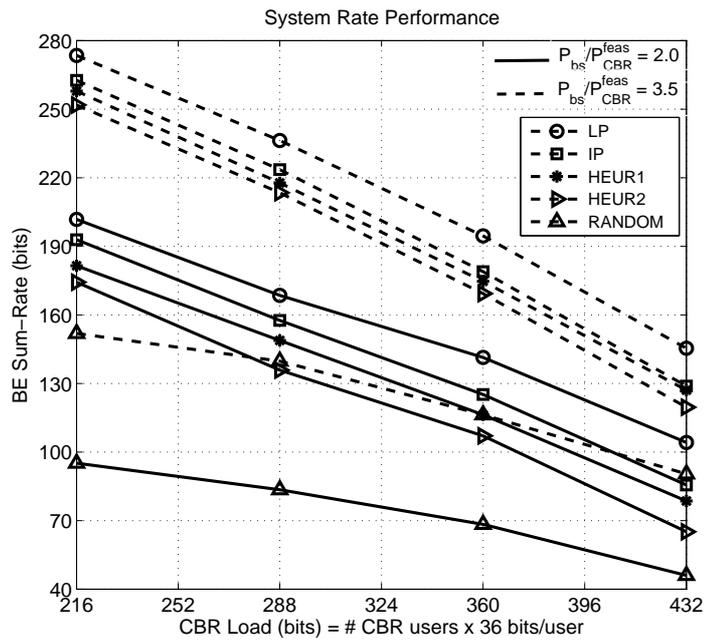}
\caption{System Rate performance vs CBR-Loading conditions}
\label{fig:Thr-Load}
\end{figure}

\begin{figure}
\centering
\includegraphics[width=0.75\textwidth]{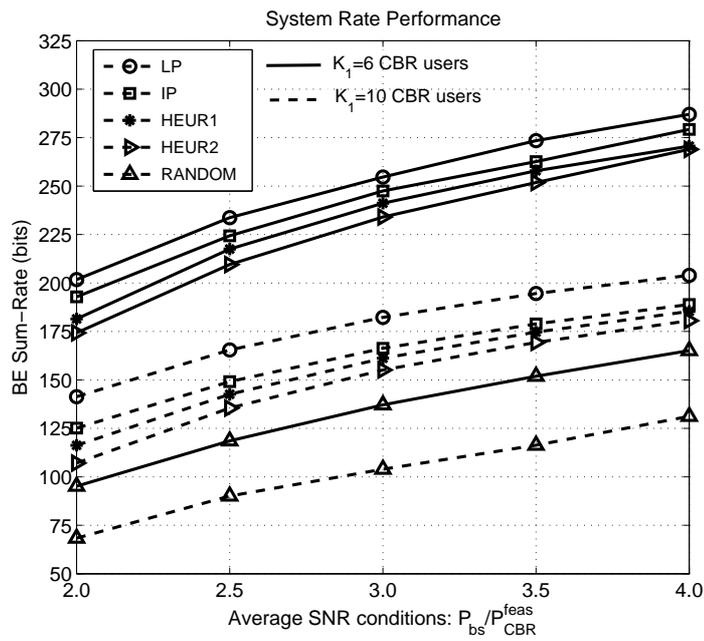}
\caption{System Rate performance vs average received SNR conditions}
\label{fig:Thr-Power}
\end{figure}

\begin{figure}
\centering
\includegraphics[width=0.78\textwidth]{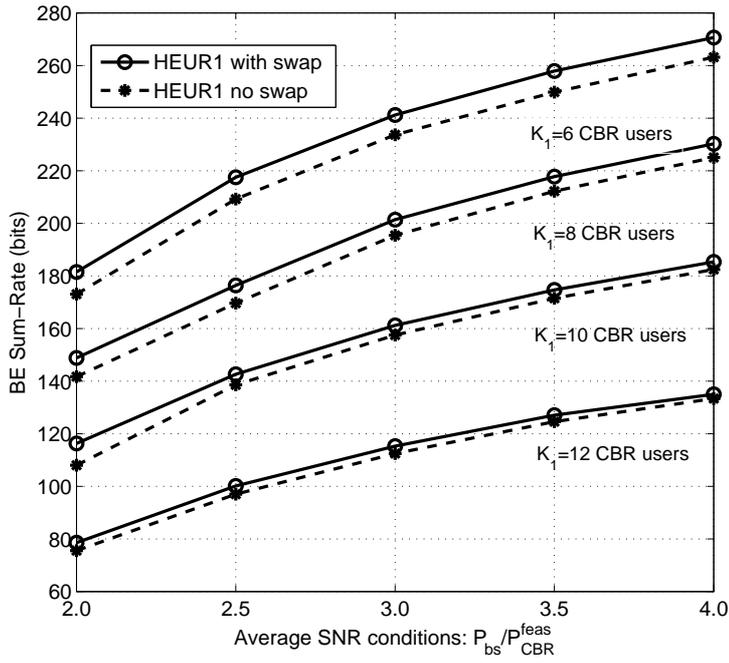}
\caption{The effect of the subchannels swap procedure to the HEUR1 performance}
\label{fig:HEUR1SwapEffect}
\end{figure}

\begin{figure}
\subfigure[Lower SNR-conditions]{
\includegraphics[scale=0.52]{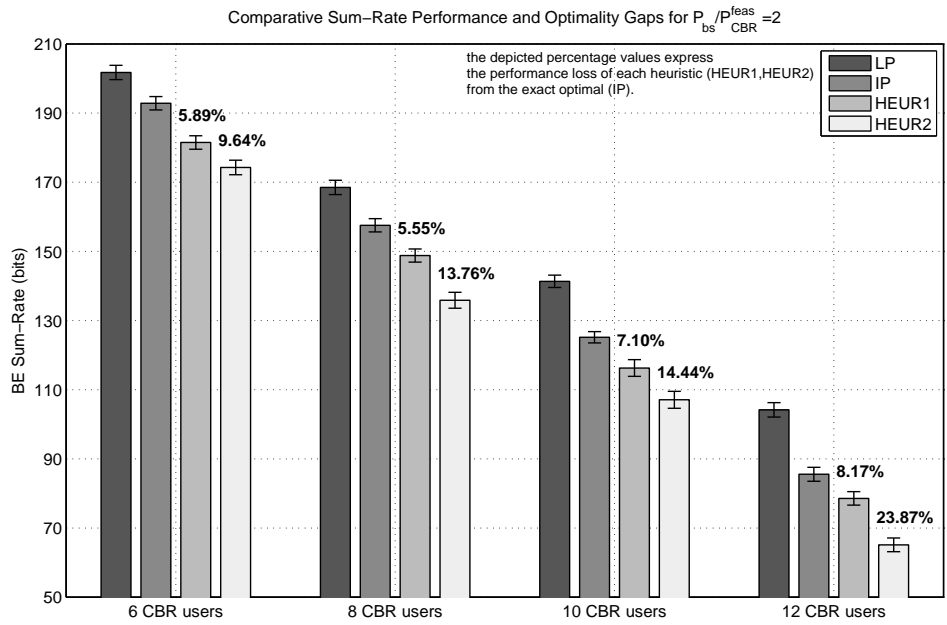}
\label{fig:Thr-Load-Bars-LowerSNR} } \subfigure[Higher SNR-conditions]{
\includegraphics[scale=0.52]{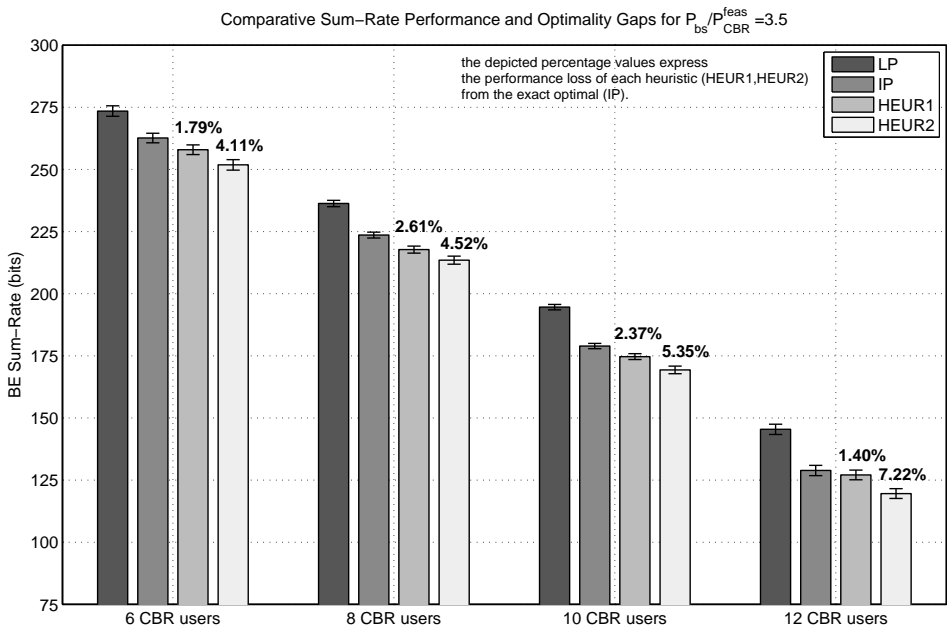}
\label{fig:Thr-Load-Bars-HigherSNR} } \caption[]{Sum-Rate and Optimality Gap
Performance}
\end{figure}

\paragraph{Results and Discussion}
In Figs.\ref{fig:Thr-Load} and \ref{fig:Thr-Power} we depict the sum-rate
performance for all schemes as a function of the required CBR loading and the
average experienced SNR conditions. The effect of the subchannel swapping procedure to the performance of the HEUR1 algorithm is illustrated in Fig.\ref{fig:HEUR1SwapEffect} (HEUR1 with/no swap). In Figs.\ref{fig:Thr-Load-Bars-LowerSNR} and
\ref{fig:Thr-Load-Bars-HigherSNR} we provide the optimality gaps of our heuristic schemes (HEUR1,HEUR2) for two representative power availability
scenarios. Finally in Table~\ref{ta:AllScenPerf} we present the achieved performance of the
heuristic schemes  compared to the exact optimal (IP) as well as
the performance difference between the IP and the relaxed LP approach. The main
observations/comments regarding the behavior of the various schemes/algorithms
are the following.

\begin{itemize}
  \item [(i)] Both heuristic approaches perform close to the optimal allocation scheme
  and follow its performance trend for different system conditions (Figs.\ref{fig:Thr-Load},\ref{fig:Thr-Power}) while
  their complexity cost is significantly lower.
  \vspace{3pt}

  \item [(ii)] The proposed heuristics outperform the RANDOM assignment scheme significantly. This means that exploiting the inherent 2D rate selectivity/diversity of the system leads to a dramatic increase on the cell performance. The performance gain of the 1st heuristic is 60.6\% on average. In particular in low SNR conditions the gain is approximately 80\% and at higher SNR conditions drops to 48.64\%, which is still high. This is reasonable, since at deteriorating channel conditions, the selection of the ``best" subchannels-set for each user becomes more critical. On the other hand the 2nd heuristic provides us with an average gain of 52.8\%, which is still remarkable.
      \vspace{3pt}

  \item [(iii)] The optimality gap (defined as the percentage sum-rate loss from the optimal scheme)
  for each algorithm depends on the
  CBR-loading and the received SNR conditions experienced in the cell. The gap
  seems to narrow as SNR conditions are improved through
  the increase in the BS power. This is justified by the fact that for higher
  amounts of BS power, larger data-rates per subchannel are supported (close to the upper bound of 6 bits), leading to the vast majority of available bandwidth resources experiencing high performance, and thus the selection
  of the optimal subchannel set for each user is not so critical anymore.
  \vspace{3pt}

  \item [(iv)] The 1st heuristic has an average optimality gap of 6.7\% for
  the lowest BS-power scenario which decreases to 2.3\% for the highest BS-power scenario.
  On the other hand the alternative dual heuristic performs worse than the first
  one. Its recorded optimality gap is 15.4\% for the lowest SNR scenario and
  4.6\% for the highest on average.
  \vspace{3pt}

  \item [(v)] The worst-case performance for the 2nd heuristic is observed at
  the lowest SNR and highest CBR-loading scenario, where the performance loss compared to
  the optimal scheme is 23.87\% (Fig.\ref{fig:Thr-Load-Bars-LowerSNR}--rightmost group of bars).
  Recall that the particular algorithm first
  allocates the available resources ignoring the demanded QoS levels. Under
  such conditions, the CBR-class sum-rate heavily dominates the overall dual-class sum-rate, and thus the subsequent resources reallocation phase finds great
  difficulty in leading to a feasible allocation. In other words when the propagation conditions are harsh and the QoS targets demanding, it is better to first guarantee the strict CBR rate constraints and then look for an improved allocation. On the contrary at higher
  SNR/lower CBR-loading conditions (Fig.\ref{fig:Thr-Load-Bars-HigherSNR}--leftmost group of bars) the achieved performance is quite high (or
  equivalently the associated optimality gap is quite low) since the distance between the original solution and the feasibility is
  significantly smaller.
  \vspace{3pt}

  \item [(vi)] The performance of the 1st heuristic is not significantly
  affected by the CBR-loading conditions, contrary to the 2nd heuristic
  behavior. For the lowest power scenario the optimality gap of the 1st heuristic increases by
  2.3\% (from 5.9\% goes to 8.2\%)
  as the number of CBR users increases from 6 to 12, whereas for the 2nd
  heuristic the corresponding increase is 14.3\% (from 9.6\% goes to
  23.9\%). This is justified by the fact that the 1st heuristic focuses on
  finding a feasible solution by prioritizing CBR-class users at the
  initial allocation phase. Similar conclusions may be drawn for higher SNR scenarios as well.
  \vspace{3pt}

  \item[(vii)] As far as the 1st heuristic the importance of the subchannel swapping step is demonstrated through Fig.\ref{fig:HEUR1SwapEffect}, where one may observe the improvement level of the sum-rate performance for all the simulation experiments. The performance gain varies from 1.2--7.7\% and it is more noticeable in lower CBR-loading scenarios. In such cases a lower number of subchannels is required for the CBR-class and thus a larger number of re-asssignments is expected to occur.
    \vspace{3pt}

  \item [(viii)] Averaged over all conditions, the 1st
  heuristic achieves 96.21\% of the optimal sum-rate performance while the 2nd 91.63\% of
  it.
  \vspace{3pt}

  \item [(ix)] The performance gap between the actual optimal solution (IP) and
  the upper bound (LP) is not negligible. For the lowest power scenario (Fig.\ref{fig:Thr-Load-Bars-LowerSNR}) this gap
  ranges between 4.4\% and 17.9\% whereas for higher BS power (Fig.\ref{fig:Thr-Load-Bars-HigherSNR}) varies between 2.8\%
  and
  11.9\%. Therefore, although the LP solution is extracted very efficiently
  compared to the IP, it often fails on providing a tight bound on the exact optimal
  performance. Consequently, if the LP bound is used as a performance benchmark
  for the evaluation of a suboptimal scheme, then the actual efficiency of the
  latter would be underestimated.
\vspace{9pt}

\end{itemize}

\section{Conclusions - Future Work}\label{sec:6}
The resource allocation problem of maximizing the sum-rate performance for a
downlink OFDMA single-cell network (like LTE) assuming heterogeneous traffic
requests and uniform power loading over the system subchannels was studied in
this work. The problem was first mathematically modeled as an ILP optimization
problem, allowing us to extract the actual optimal allocation decision and the
associated maximum achieved sum-rate performance in reasonable execution time.
By relaxing the integrality constraints on the above model, a performance upper
bound may also be extracted with  3rd order polynomial complexity cost. In the
second part of this work, motivated by the resemblance of the specific problem
with a well-known combinatorial optimization problem that is, the Generalized
Assignment, we developed two heuristic algorithms for efficiently allocating
system subchannels to the competing classes and users. We finally demonstrated
through extensive simulation experiments that the performance loss of the
heuristic schemes compared to the optimal is rather low, especially for the 1st
heuristic, and that our schemes heavily outperform semi-random subchannel assignments. Possible
extensions of this work may involve the cooperation of the proposed schemes
with time-domain scheduling algorithms handling packet delay and fairness QoS objectives, the employment of practical channel
state reporting schemes and the consideration of interference in multi-cell
deployments.

\begin{table}[!h]
\centering \footnotesize \caption{Achieved Optimality Gap values for all
Scenarios} \vspace{6pt} \label{ta:AllScenPerf}
\begin{tabular}{llll}
\hline\noalign{\smallskip}
\textbf{Scenario} & \textbf{IP/LP} & \textbf{HEUR1/IP}   & \textbf{HEUR2/IP}\\
\text{\{${K_1,P_{bs}/P_{\text{CBR}}^{\text{feas}}}$\}} & \% & \% & \%\\[2pt]
\hline\\[1pt]
\{6, 2.0\}  & 95.59 & 94.11 & 90.36 \\
\{6, 2.5\}  & 95.99 & 96.94 & 93.41 \\
\{6, 3.0\}  & 97.18 & 97.46 & 94.58 \\
\{6, 3.5\}  & 96.03 & 98.21 & 95.89 \\
\{6, 4.0\}  & 97.29 & 96.92 & 96.32 \\
\{8, 2.0\}  & 93.50 & 94.45 & 86.24 \\
\{8, 2.5\}  & 93.60 & 93.77 & 92.16 \\
\{8, 3.0\}  & 94.11 & 96.74 & 94.31 \\
\{8, 3.5\}  & 94.62 & 97.39 & 95.48 \\
\{8, 4.0\}  & 94.91 & 97.80 & 96.23 \\
\{10, 2.0\} & 88.55 & 92.90 & 85.56 \\
\{10, 2.5\} & 90.12 & 95.61 & 90.83 \\
\{10, 3.0\} & 91.24 & 96.93 & 93.28 \\
\{10, 3.5\} & 91.94 & 97.63 & 94.65 \\
\{10, 4.0\} & 92.62 & 98.11 & 95.59 \\
\{12, 2.0\} & 82.12 & 91.84 & 76.12 \\
\{12, 2.5\} & 85.54 & 94.51 & 85.27 \\
\{12, 3.0\} & 88.09 & 96.44 & 90.05 \\
\{12, 3.5\} & 88.62 & 98.60 & 92.78 \\
\{12, 4.0\} & 91.11 & 97.97 & 95.40 \\
\noalign{\smallskip}\hline
\end{tabular}
\end{table}

\appendix
\section{The Data-Rate Abstraction Model}\label{sec:app1} Let
$\left\langle {n,k} \right\rangle$ an arbitrary subchannel/user combination and
assume that a bit-stream is transmitted from the BS to the $k^{th}$ user over
the $n^{th}$ subchannel. We denote by $P_{n,k}$ the allocated transmitted
power, ${\left| {{h_{n,k}}} \right|^2}$ the propagation channel power gain
(which is known at both transceiver ends), $N_0$ the noise power density and
$\Delta f$ the bandwidth of each subchannel. Then, by applying the closed-form
approximation model of~\cite{Pi06}, the achieved data-rate (channel capacity)
for preserving a minimum transmission error rate of $P_e$ will be given by
Eq.\eqref{eq:RateCalc}, where $f$ stands for the abstraction function.

\begin{equation}\label{eq:RateCalc}
{r_{n,k}} = f\left( {{P_{n,k}},\left| {{h_{n,k}}} \right|,{P_e}} \right) =
{\log _2}\left( {1 + \frac{{{P_{n,k}} \cdot {{\left| {{h_{n,k}}} \right|}^2}}}
{{\frac{1} {3}{{\left[ {{Q^{ - 1}}\left( {\frac{{{P_e}}} {4}} \right)}
\right]}^2} \cdot {N_0} \cdot \Delta f}}} \right)
\end{equation}

We further define the normalized channel-to-noise ratio as in
Eq.\eqref{eq:channTonoise}, and by employing the Uniform Power Loading
assumption, we resort to the expression of Eq.\eqref{eq:RateCalc_Final} where
$P_{bs}$ is the total available BS power. As also seen in
Eq.\eqref{eq:RateCalc_Final} the achieved data rate on each subchannel is
hard-limited by the highest available transmission order denoted by $c_{max}$
bits. The latter comprises a system constraint similar to the BS power.
Finally, notice that if power loading is not a-priori known then the achieved
data-rates can not be precalculated, since they depend on the allocated amount
of power. In such scenarios joint power and subchannel allocation must be
employed (see~\cite{GoKo10} for example).

\begin{equation}\label{eq:channTonoise}
{\gamma _{n,k}} = {{{{\left| {{h_{n,k}}} \right|}^2}} \mathord{\left/
 {\vphantom {{{{\left| {{h_{n,k}}} \right|}^2}} {\left[ {\left( {{1 \mathord{\left/
 {\vphantom {1 3}} \right.
 \kern-\nulldelimiterspace} 3}} \right) \cdot {{\left[ {{Q^{ - 1}}\left( {{{{P_e}} \mathord{\left/
 {\vphantom {{{P_e}} 4}} \right.
 \kern-\nulldelimiterspace} 4}} \right)} \right]}^2} \cdot {N_0} \cdot \Delta f} \right]}}} \right.
 \kern-\nulldelimiterspace} {\left[ {\left( {{1 \mathord{\left/
 {\vphantom {1 3}} \right.
 \kern-\nulldelimiterspace} 3}} \right) \cdot {{\left[ {{Q^{ - 1}}\left( {{{{P_e}} \mathord{\left/
 {\vphantom {{{P_e}} 4}} \right.
 \kern-\nulldelimiterspace} 4}} \right)} \right]}^2} \cdot {N_0} \cdot \Delta f}
 \right]}}
\end{equation}

\begin{equation}\label{eq:RateCalc_Final}
{r_{n,k}} = \min \left\{ {{{\log }_2}\left( {1 + \frac{{{P_{bs}}}} {N} \cdot
{\gamma _{n,k}}} \right),{c_{\max }}} \right\}
\end{equation}

\bibliographystyle{IEEEtran}
\bibliography{IEEEfull,gotsis-STSbib-REV}

\end{document}